\documentclass[conference]{IEEEtran}
\IEEEoverridecommandlockouts

\RequirePackage{etex}
\usepackage[pdftex]{graphicx}
\graphicspath{{../pdf/}{../jpeg/}}
\DeclareGraphicsExtensions{.pdf,.jpeg,.png}

\usepackage[cmex10]{amsmath}
\usepackage{algorithmic}
\usepackage{array}
\usepackage{mdwmath}
\usepackage{mdwtab}
\usepackage{eqparbox}
\usepackage{url}
\usepackage{tikz}
\usetikzlibrary{matrix,chains,positioning,decorations.pathreplacing,arrows}
\usepackage{mathtools}
\usepackage{tikz-network}

\usepackage[hidelinks]{hyperref}
\hypersetup{
    colorlinks=false,
    linkcolor=blue,
    filecolor=magenta,      
    urlcolor=cyan,
}
\usepackage{cite}
\usepackage{amsmath,amssymb,amsfonts}
\usepackage{algorithmic}
\usepackage{graphicx}
\usepackage{textcomp}
\usepackage{xcolor}
\def\BibTeX{{\rm B\kern-.05em{\sc i\kern-.025em b}\kern-.08em
    T\kern-.1667em\lower.7ex\hbox{E}\kern-.125emX}}

\begin{document}

\onecolumn

\noindent \textcopyright{} 2019 IEEE. Personal use of this material is permitted. Permission from IEEE must be obtained for all
other uses, in any current or future media, including reprinting/republishing this material for advertising or
promotional purposes, creating new collective works, for resale or redistribution to servers or lists, or reuse
of any copyrighted component of this work in other works.

\twocolumn

\title{Modeling Gate-Level Abstraction Hierarchy Using Graph Convolutional Neural Networks to Predict Functional De-Rating Factors \\
\thanks{This work was supported by the RESCUE ETN project. The RESCUE ETN project has received funding from the European Union's Horizon 2020 Programme under the Marie Skłodowska-Curie actions for research, technological development and demonstration, under grant No. 722325}
}

\author{%
\IEEEauthorblockN{%
  Aneesh Balakrishnan\IEEEauthorrefmark{1}\IEEEauthorrefmark{2},
  Thomas Lange\IEEEauthorrefmark{1}\IEEEauthorrefmark{3},
  Maximilien Glorieux\IEEEauthorrefmark{1},
  Dan Alexandrescu\IEEEauthorrefmark{1},
  Maksim Jenihhin\IEEEauthorrefmark{2}%
}
\IEEEauthorblockA{%
  \IEEEauthorrefmark{1}\textit{iRoC Technologies}, Grenoble, France \\
   \IEEEauthorrefmark{2}\textit{Department of Computer Systems, Tallinn University of Technology}, Tallinn, Estonia \\
  \IEEEauthorrefmark{3}\textit{Dipartimento di Informatica e Automatica, Politecnico di Torino}, Torino, Italy \\
  \{aneesh.balakrishnan, thomas.lange, maximilien.glorieux, dan.alexandrescu\}@iroctech.com \qquad
  maksim.jenihhin@taltech.ee}
}

\maketitle

\begin{abstract}

The paper is proposing a methodology for modeling
a gate-level netlist using a Graph Convolutional Network (GCN).
The model predicts the overall functional de-rating factors of
sequential elements of a given circuit. In the preliminary phase
of the work, the important goal is making a GCN which able to
take a gate-level netlist as input information after transforming
it into the Probabilistic Bayesian Graph in the form of Graph
Modeling Language (GML). This part enables the GCN to learn
the structural information of netlist in graph domains. In the
second phase of the work, the modeled GCN trained with a
functional de-rating factor of a very low number of individual
sequential elements (flip-flops). The third phase includes the
understanding of GCN model’s accuracy to model an arbitrary
circuit netlist. The designed model validated for two circuits.
One is the IEEE 754 standard double precision floating point
adder and the second one is the 10-Gigabit Ethernet MAC IEEE
802.3 standard. The predicted results compared to the standard fault
injection campaign results of the error called Single Event Upset
(SEU). The validated results are graphically pictured in the form
of the histogram and sorted probabilities and evaluated with
the Confidence Interval (CI) metric between the predicted and
simulated fault injection results.

\end{abstract}



\section{Introduction}

System engineering advances and focusing on the integration of small-scale technologies in the system building process. The realization of full potential micro- and nanoscale devices highlights the challenges faced by electronics businesses industries in maintaining or improving their technological competitiveness. System engineering and its challenges keep the design engineers more concentrating on the reliability issues with their designed systems. Focusing on the reliability problems occurring with micro- and nanoscale technology development and its impact on everything from the design phase to actualized products in the health, automotive, aerospace, communication, and many other fields, the system design engineers considering all possible methodological precautions to prevent reliability issues based on their criticality. The industrial customers demanding high-quality reliable devices and in order to meet the requirements, research and design departments proposing different metrics which ensures a default standard quality and reliability.  One of the major threats in the system's reliability is the Single Event Effects (SEE) due to the cosmic rays and electromagnetic radiation. Cosmic rays are particles that hit the Earth's atmosphere from space. They include protons, helium nuclei (like $\alpha$ radiation), and electrons (like $\beta$ radiation). The radiations like gamma and X rays, which are electromagnetic and indirectly ionizing radiations.  The two major consequences of the SEE are Single Event Transients (SET) and Single Event Upset (SEU). The effect of SEU and SET at the functional level of the circuit is known as functional de-rating factors. They are more closely examined here with help of exhaustive and accelerated fault injection campaigns. 

The important aspect of the fault injection campaign is the more reliable and accurate information over other different mathematical models. In contrary to this point, the effort in terms of time is non-feasible from the perspective of a designer. The effort in the non-feasible dimension of work can be reduced by implementing different statistical and mathematical models. This research goal achieving through the proposed model and automate the assessment of different reliability factors within feasible time constraints and making a trade-off with accuracy.

\subsection{Motivation}

Even though above-explained networks are eligible to do deep learning, the implemented model predominantly depending on the neural network as referenced in \cite{Franco} and \cite{William}. The cited paper \cite{Franco} extends the current neural networks to a new model called Graph Neural Network (GNN) and process the data in graph domains. There is a lot of scientific areas of engineering which deals the information in the graph domains. This point is considered to be a decisive moment of the thought towards a representation of the gate-level circuit information in graph domains and feeding to a graph neural network for processing it. Here, the implemented model adopts a form of Graph Convolutional Network (GCN) proposed by Thomas.N.Kipf in \cite{kipf2017semi}, which is another version of generally called graph neural networks. This model is particularly briefed in section IV and V. 

\subsection{Organization of the Paper}

The introduced paper includes nine sections in total. Section I generalizes the facts and issues in the field of reliability engineering and followed by the main motivation of this work as well as the organizational structure of the paper. Section II gives a background introduction to neural networks and different reliability factors of the micro-electronic systems. This part dedicated to explaining the background of this work. In Section III, the main methodological implementation overview is given, whereas in section IV and V, GCN model and it's neuron implementation explained with mathematical equations. That is, sections IV and V together constitutes the methodology and model architecture and their in-depth view. Section VI illustrates the results and their validations in terms of 95 \% Confidence Interval (CI), histograms and sorted order of FDR probabilities. Section VII describes the main model drawbacks. Future works and their importance with their probability to achieve, are discussed in section VIII and  In Section IX, the whole work and it's holistic approaches are concluded.


\section{Background}
\subsection{Interpretation of Standard Electrical Terms}
\subsubsection{\textbf{SEE Analysis Concepts}} 
As the term suggests, a single event effect (SEE) results as the penetration consequences of the energetic radiation particle. The main consequence effects are classified as two categories destructive and non-destructive. The Single Event Upset (SEU) and Single Event Transient (SET) are considered to be non-destructive and soft-errors. The radiation hazards like Single Event Latchup (SEL) are categorized under hard-errors (or) destructive-type faults. An elaborate explanation for different radiation hazards can be referenced from \cite{Soft:Errors}. This work is mainly contributing to the derating analysis of Single Event Upset in sequential elements. The quantitative analysis of SEE is based on different derating factors, called Functional derating, Logical derating, Temporal derating, and Electrical derating. 

\subsubsection{\textbf{Electrical Derating}}
    The Electrical Derating (EDR) evaluating the effect of modeled logic SET pulse that has the same effect in the circuit as the original analog SET pulse. SET pulse can be modeled logically as an inversion of the output signal amplitude of combinational cells in gate-level abstraction. The effect of such types of anomalies with various electrical factors like electrical pulse width and electrical amplitude range defines how well a transient error obstructs the standard signal propagation in the given circuit.  
    
\subsubsection{\textbf{Temporal Derating}}
    Temporal Derating (or) Time derating associate to the opportunity window ascribed to SEE error (SET (or) SEU) and it's probability to be latched to the downstream sequential elements like Flip-Flop, Latch and Memory.   

\subsubsection{\textbf{Logical Derating}}
    The logical vulnerability of the SEU within the combinational cell network based on their logic functions is quantified with masking effect probability, termed as Logical Derating (LDR) factor.  
    
\subsubsection{\textbf{Functional Derating}}
    Functional Derating evaluates how likely the soft error propagate to make an observable impact on the functioning of the circuits or systems.

\subsection{The reasoning of Graph Convolutional Neural Network}

\begin{figure}[ht]
    \centering
    \begin{tikzpicture}

        \draw[fill=gray!30!white] (0,0) ellipse (90pt and 100pt);
        \draw[fill=gray!70!white] (0,-1.05) ellipse (60pt and 70pt);
        \draw[fill=gray!100!white] (0,-1.75) ellipse (40pt and 50pt);
    
        \draw (0,2) node[circle] {Artificial Intelligence};
        \draw (0,0.5) node[circle] {Machine Learning};
        \draw (0,-1.3) node[circle] {Deep Learning};

        
    \end{tikzpicture}
    \caption{A relational analysis of Artificial Intelligence (AI), Machine Learning (MI) and Deep Learning (DL)}
    \label{Figure_AI}
\end{figure}
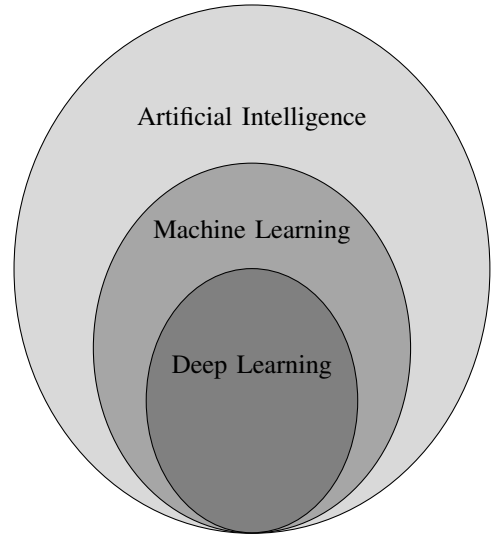

The part actually gives a clear idea of the relationship between Artificial Intelligence (AI), Machine Learning (ML) and Deep Learning (DL) to the reader. The main point of view is, Deep Learning or Deep Neural Networks (DNN) or Artificial Neural Networks (ANN) are commonly considered as a subset of machine learning which in turns derived from the concept of artificial intelligence. The machine learning in which data parsed for learning phase and then apply the learned dependencies of the data features to arrive at a decision, whereas, in case of deep learning algorithms, it appears in layers to create an Artificial Neural Network (ANN) that can learn and make intelligent decisions on its own. The artificial intelligence considered to be a global idea of ML and DL and it can be defined by a way of enabling the machine (e.g.\ a computer) to attain a given task based on a stipulated set of rules called an algorithm.

As mentioned in the above paragraph, the deep neural networks are able to make intelligent decisions on its own, the work which experimented here mainly based on a neural network, called Graph Convolutional Neural Network. Intelligent network like GCN is actually different from traditional neural networks algorithm and slightly varied from traditional Convolutional Neural Networks (CNN). A normal neural network consists of staked hidden layers, where each of the neurons (or) nodes from the current layer receives input from all the nodes from the previous layer, commonly known as dense layers. Then performs a dot product of the data at the input of the neuron and the weights of the neuron and passed through an activation function respectively. These determined values passing to the successive layers by concatenating the input, hidden and output layers together. CNN is different from the traditional way of constructing the dense layered neural network. In CNN, the initial input features are convolved with kernel input filters and then down-sampled through a pooling layer and finally directed to a normal fully connected neural network.


\section{Overview of the Work}

\begin{figure}[ht]
    \includegraphics[width=\linewidth]{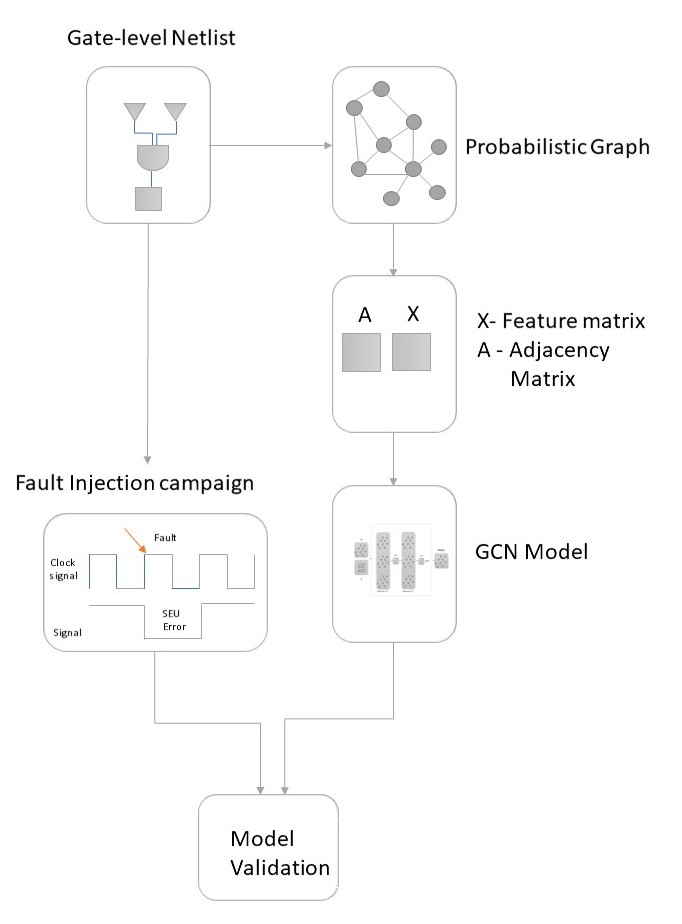}
    \caption{Systematic block diagram of the scientific work }
    \label{figure_block}
\end{figure}

A better overview of the work portrayed in figure \ref{figure_block}.
Before stepping into the detailed structure of the whole work, it is very relevant to brief the importance of mapping the gate level netlist into the probabilistic graph model. The more the mapping achieve accuracy, the more the model delivers a valid result because the graph structure maintains the required statical information. In order to execute the preliminary aim of the work, different user-defined Verilog Procedural Interface (VPI) functions had written and it in turn applied to extract all the relevant details of the gate level netlist and formatted into a probabilistic graph model through GML attributes. As stepping forward into the successor stage of the work, GCN adopted as the model in order to learn the whole designed probabilistic graph. The more comprehensively explained hierarchical architecture of GCN updated in the successive sections.  

The netlist representation in graph domains subsequently used to extract the adjacency matrix, which represented by A in the figure \ref{figure_block}. Correspondingly, a feature matrix X also obtained by the random walk method using the node2vec algorithm. The random walk method gives a feature vector corresponding to a node with respect to its neighboring nodes. The feature vector is mainly based on transition probabilities from source to target nodes and also the degree of nodes.  

These are the two main inputs given to the GCN model. GCN then commenced learning the whole netlist as a probabilistic graph. As soon as, it processes the adjacency matrix and feature matrix, a model of the netlist is delivered. After that, this model is used for the training phase and testing phase for accomplishing the FDR prediction goal. Finally, the predicted data is compared with the fault injection campaign FDR data.

The whole deep learning framework was implemented in MXNet. 


\section{Graph Convolutional Network}

\subsection{Recent Literature History}

Different prodigious research work had been introduced in the past decades of years, for generalizing the conventionally established neural network like Recurrent Neural Network (RNN) or Convolutional Neural Network (CNN) for working on arbitrarily structured graphs, even though it is a great challenging problem. 

This work is mainly based on the GCN neural network. A similar spectral approach introduced in \cite{Michal}. By the GCN model, it is able to exemplify the spectral rule approach in the graphical learning process and it achieves significantly faster training times with higher predictive accuracy and also reaching state-of-the-art classification results on a number of benchmark graph datasets. 

\subsection{Architecture}

\begin{figure}[ht!]
    \includegraphics[width=\linewidth]{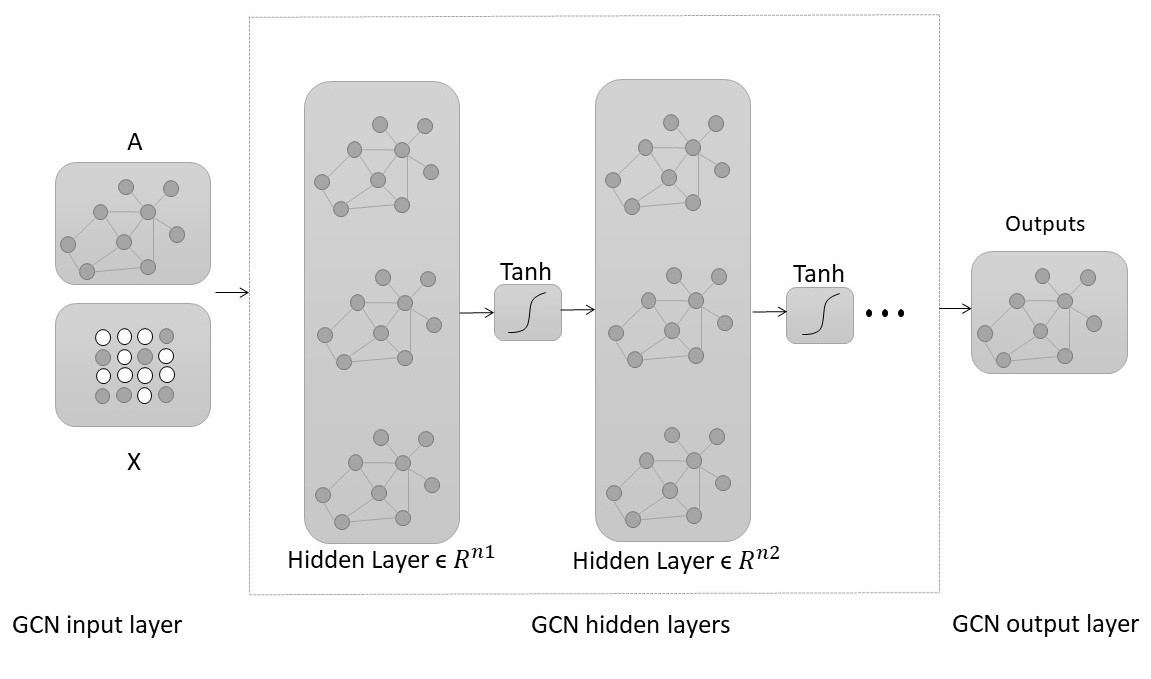}
    \caption{GCN model \cite{kipf2017semi}}
    \label{gcn_architecture}
\end{figure}

Figure \ref{gcn_architecture} provide a architectural view of GCN. The work made a GCN model of two hidden layers as given in figure \ref{gcn_architecture}. The first layer in this work contains 4 hidden nodes and the second layer contains 2 hidden nodes. These two hidden layers stacked between the input layer and the output layer. The input layer contains a number of nodes which equivalent to the gate-level netlist elements of the circuits. It varies from circuit to circuit. The model can able to model even for a  large number of elements of the circuit by this time. But it is difficult to say a limit now. Both hidden layer's nodes activated by the non-linear function called a hyperbolic tangential function (Tanh). During the training phase, the model is updating at each step and optimized by an adaptive learning rate optimization algorithm called 'Adam' \cite{Kingma:2014vow}. The dimension of the hidden layers can be chosen by arbitrarily and it depends on the parsed adjacency matrix.

\subsection{Model}

The Graph Convolutional Network is a powerful neural network architecture for machine learning on graphs. Following paper \cite{kipf2017semi}, revealing the fact that most of the graph neural networks has been addressing a common architecture in general, which lead to the name called Graph Convolutional Neural Networks (GCN). The convolution name comes after using the filter parameters shared across all locations of the graph. 

\subsubsection{Model Definition}

The created probabilistic graph model of the gate-level netlist embedded into the GCN network with the intention of learning the function of features in the graph. The graph described as a $G = (\nu,\varepsilon)$, where $\nu$ represents vertices or nodes and $\varepsilon$ represents the edges between the vertices. The graph characterized as, 

{$\odot$} : Every nodes $i$ is attributed with feature vector $x_i$ of dimension $D$. So for $N$ nodes, we have feature matrix $X \colon N  \times D$.  

{$\odot$} : Another important parameter is the adjacency matrix A, which indicates the graph structure.

{$\odot$} : The propagation rule will produce a node-level output of $Z \colon N  \times F$, where the F represents a feature vector of each output node.   

{$\odot$} : Every neural network layer can be represented as in equation \ref{layer}.

\begin{equation}
\label{layer}
    H^{(l+1)} = f(H^{(l)},A)
\end{equation}

Where $H^{(l+1)}$ represents the any hidden layer node matrix at $(l+1)^{th}$ level and it equivalent to the function of previous hidden layer node matrix $H^l$ at $l^{th}$ level and the adjacency matrix A. $H$ can be taken as the feature matrix $X$ at initial level, ie $H^{(0)} = X$ and $Z$ at final level. $Z$ represents the graph level output.  

\subsubsection{Model Propagation Rule}
In this whole paper, an exact propagation model for the Graph Neural Network is adapted to tackle the prediction problem. 

A simple form of the layer-wise propagation rule abbreviated as: 

\begin{equation}
    f(H^{l},A) = \sigma(A H^l W^l)
\end{equation}

Where, $W^l$ is the $l^{th}$ neural network weight matrix and $\sigma()$ is the activation function like Rectified Linear Unit (ReLU), while this work utilizes a hyperbolic tangent activation function (Tanh). Even though the above propagation rule seems to be very simple, it was proved to be very powerful. The major disadvantage of this kind of model is the adjacency matrix A, which not normalized so that multiplication of A with feature matrix will change the scale of feature matrix completely. The second problem as mentioned by the authors of this model is, the model does not consider the self-features by a node itself. And the problem is completely taken away by providing an identity matrix for the nodes.

The major problem overcame by a normalizing matrix A. Normalization achieved by an inverse diagonal node degree matrix D, such as the rows of $D^{-1} A$ sums to 1. So the multiplication becomes more similar to taking the average of neighboring nodes. This lead to symmetric normalization i.e, $D^{-\frac{1}{2}} A D^{-\frac{1}{2}}$, and it more than just a mere averaging of neighboring features. These combined methods used in this work as a propagation rule which exactly similar to the way implemented in paper \cite{kipf2017semi} and final layer-wise propagation rule provided as:

\begin{equation}
    f(H^{l},A) = \sigma(\hat{D}^{-\frac{1}{2}} \hat{A} \hat{D}^{-\frac{1}{2}} H^{l} W^{l})
\end{equation}
where, $\hat{A} = A + I;$ with I defined as identity matrix and $\hat{D} $ is the diagonal degree node matrix of $\hat{A}$.  

\subsubsection{Input Feature Matrix}

In order to generate a feature matrix corresponding to the nodes in the probabilistic graph, we use a node2vec algorithm provided by \cite{node2vec}. \textit{node2vec} is an algorithmic framework for learning continuous feature representations of nodes in networks. According to this algorithm, it maps nodes to the low-dimensional feature space which maximizes the likelihood of preserving network neighborhoods of nodes. This objective is optimizing by the stimulated biased random walks. It preserves a spectrum of equivalences from homophily to overall structural equivalence, by anticipating a balanced exploration-exploitation trade-off. 


\section{GCN Neuron Model}

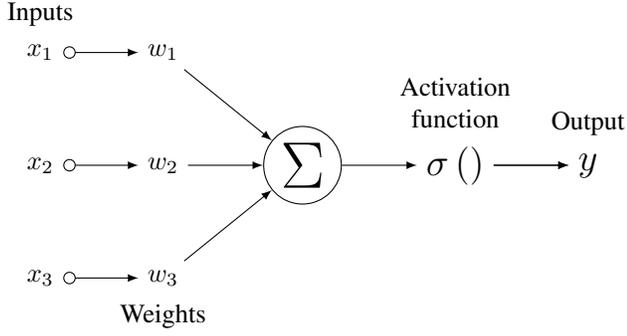
\begin{figure}[ht]
\centering

\begin{tikzpicture}[
    init/.style={ 
         draw, 
         circle, 
         inner sep=2pt,
         font=\Huge,
         join = by -latex
    },
    squa/.style={ 
        font=\Large,
        join = by -latex
    }
]


\begin{scope}[start chain=1]
    \node[on chain=1,label=above:Inputs] at (0,1.5cm)  (x1) {$x_1$};
    \node[on chain=1,join=by o-latex] (w1) {$w_1$};
\end{scope}


\begin{scope}[start chain=2]
    \node[on chain=2] (x2) {$x_2$};
    \node[on chain=2,join=by o-latex] {$w_2$};
    \node[on chain=2,init] (sigma) {$\displaystyle\Sigma$};
    \node[on chain=2,squa,label=above:{\parbox{2cm}{\centering Activation\\ function}}] {$\sigma\left( \right)$};
    \node[on chain=2,squa,label=above:Output,join=by -latex] {$y$};
\end{scope}


\begin{scope}[start chain=3]
    \node[on chain=3] at (0,-1.5cm) 
    (x3) {$x_3$};
    \node[on chain=3,label=below:Weights,join=by o-latex]
    (w3) {$w_3$};
\end{scope}




\draw[-latex] (w1) -- (sigma);
\draw[-latex] (w3) -- (sigma);


\end{tikzpicture}

\caption{ A neuron model for embedding nodes}
\label{figure_neuron}
\end{figure}

\begin{figure}[ht!]
    \centering
    \includegraphics[width=2.5 in]{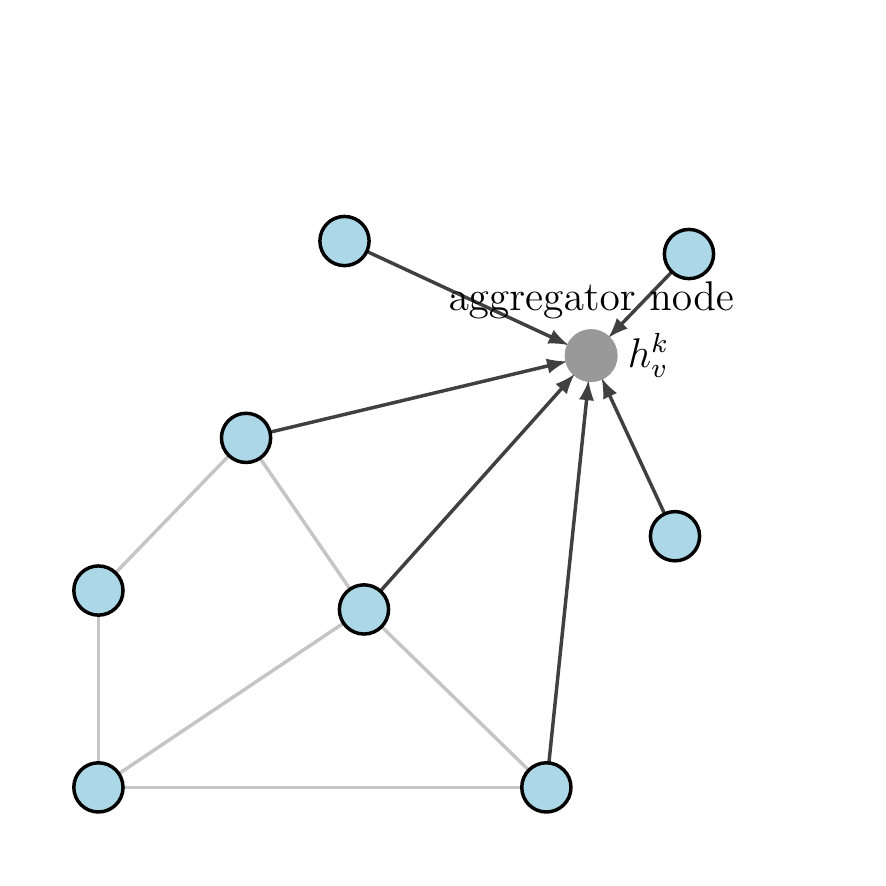}
    \caption{Aggregation model of a node}
    \label{fig:my_label_aggr}
\end{figure}

Figure \ref{figure_neuron} represents a single slice of neuron pipeline which implemented in the neural network. The GCN model neighborhood aggression is typically different from the basic neighborhood aggregation algorithm as mentioned in \ref{equ_aggr}. It is clear that to mention that, no bias factor is added and trained in the model. A similar weight matrix $W_k$ used for the self-node embedding and neighbor nodes embedding. This improves and achieves more parameter sharing across the network and down-weights the higher degree neighbors. The important thing to notice is the normalization factor which varies across the neighbors instead of a simple average. In equation \ref{equ_aggr}, the node $v$ is abbreviated for the node targeted for embedding process, while $N(u)$ in equation \ref{equ_aggr_1} represents the neighbouring nodes of $v$. $h_v^k$ given in equation \ref{equ_aggr_1} is $k^{th}$ layer node $v$ aggregator as indicated in figure \ref{fig:my_label_aggr}. The equation \ref{equ_aggr_1} pictorially represented in figure \ref{figure_neuron}. $\sigma()$ indicates a non-linear function, simply named as an activation function in figure \ref{figure_neuron}. 

\begin{align}
\label{equ_aggr}
     h_v^{0} &=X_v \\
\label{equ_aggr_1}
     h_v^{k} &= \sigma \left( W_k \displaystyle \sum_{u \epsilon N(v) \bigcup v}  \frac{h_u^{k-1}}{\sqrt{\mid N(u) \mid \mid N(v) \mid}} \right) \\
     Z_v &= h_v^{K}
\end{align}

The variable $h_v^{0}$ shows a node $v$ at the input layer and its input equivalent to the node v features vector $X_v$ extracting using a node2vec algorithm. The variable $h_v^{K}$ denotes node embedding of a node $v$ at the last layer $K$ of neural network and the output node's embedding with its features space abbreviated as $Z_v$.

Figure \ref{fig:my_label_aggr} shows an aggregator node in the network which collects related information and features of neighboring nodes.


\section{Result : Modeling and Validations}
 
As mentioned earlier in the paper, the model tested with two circuits. The very first one is the double precision floating point adder which extracted from the double precision floating point core as a submodule, which meets the IEEE 754 standard and available in the OpenCores website. The second circuit is also accessible from OpenCores as 10-Gigabit Ethernet project, where Management Data Input/Output (MDIO) function of this module designed to meet 10-Gigabit Ethernet IEEE 802.3 standard. In  MAC design based on the Xilinx LogiCORE 10-Gigabit Ethernet MAC, the transmitter and the receiver incorporate the reconciliation layer. Therefore the receive engine, as well as transmit engine, will be specifically designed to interface the client and the physical layer. 

\subsection{ Double precision floating point adder }

\begin{figure}[ht!] 
    \centering
    \includegraphics[width=\linewidth]{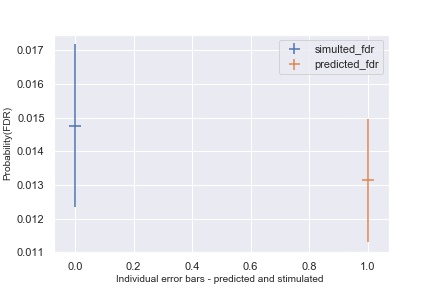}
    \caption{ An overall FDR confidence interval (CI) comparison between predicted and stimulated data}
    \label{figure_fpu_ci}
\end{figure}

Figure \ref{figure_fpu_ci} actually represents the confidence interval comparison of the predicted Functional Derating (FDR) data of flip-flops with the FDR data generated from random fault injection campaign. The CI calculated in python, by finding the mean of the flip-flop's FDR distribution and their FDR distribution error for 95\% confidence interval. There are no electrical features extracted from the circuit's gate-level netlist to train the upholding neural network model. The training had done with less than 10 flip flops FDR. The overall comparison indicates the prediction almost following the stimulated SEU fault's FDR data.  

\begin{figure}[ht!]
  \centering
  \includegraphics[width=\linewidth]{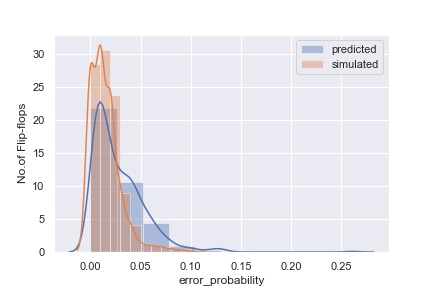}
  \caption{Histogram Comparison}
  \label{figure_fpu_hist}
\end{figure}

As observed from the histogram graph depicted in figure \ref{figure_fpu_hist}, the prediction of the FDR probability distribution function (PDF) of the flip-flops comparatively very close to the original PDF of the flip-flops.   

\begin{figure}[ht!]
  \centering
  \includegraphics[width=\linewidth]{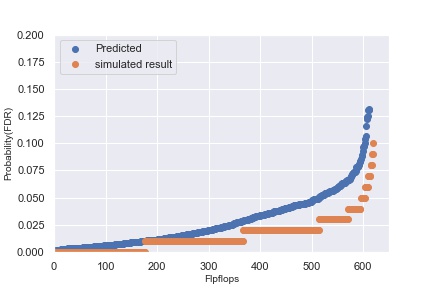}
  \caption{Sorted FDR probability graph }
  \label{figure_sorted}
\end{figure}

\begin{figure}[ht!]
\centering
\includegraphics[width=\linewidth]{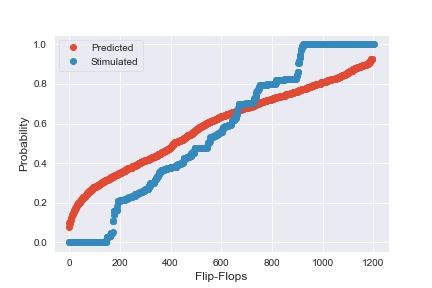}
\caption{Sorted FDR probability comparison }
\label{figure_sorted_MAC}
\end{figure}

Figure \ref{figure_sorted} compares the sorted FDR value of simulated and predicted data. This sorted FDR plot only shows how an overall functional derating curve behaves with respect to flip flops. In fact, it does not provide any individual flip-flip comparison. The work is currently extending to do that. This comparison plotted to provide an intuition to the reader that the model is actually able to get a reasonable approximation with respect to their independent structural information. This is specifically understandable when the sorted FDR graph in figure \ref{figure_sorted_MAC} from Ethernet MAC compared here, which entirely different from floating point adder.

\subsection{Ethernet MAC}

\begin{figure}[ht!]
  \centering
  \includegraphics[width=\linewidth]{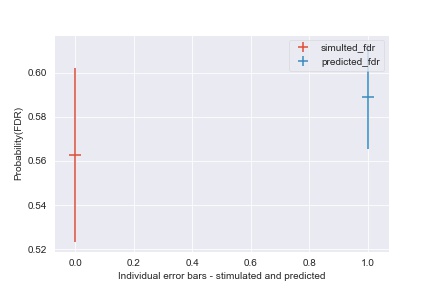}
  \caption{Representation of CI comparison }
  \label{figure_CI_MAC}
\end{figure}

\begin{figure}[ht!]
  \centering
  \includegraphics[width=\linewidth]{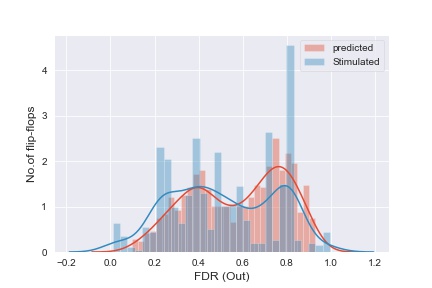}
  \caption{Representation of Histogram comparison (Filetring out some outliers) }
    \label{figure_hist_MAC_filtered}
\end{figure}

\begin{figure}[ht!]
\centering
\includegraphics[width=\linewidth]{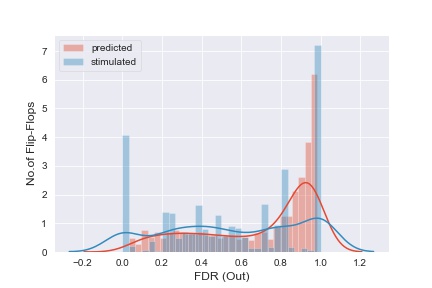}
\caption{Representation of Histogram comparison without filtering }
\label{figure_hist_MAC}
\end{figure}

Here the modeling tries to validate on Ethernet MAC circuit. This reveals how powerful is this algorithm to predict on the completely different histogram with a training sequence of 5 flip-flops (ie, less than 1 \% of overall flip-flop number). Figure \ref{figure_CI_MAC} represents the overall confidence interval comparison of the predicted FDR data with FDR data obtained from fault injection campaign on the sequential elements in each clock-cycle independently. Figure \ref{figure_hist_MAC_filtered} represents the PDF where some of the data points are filtered, which considered being outliers within the data space and plotted the remaining data. Simultaneously figure \ref{figure_hist_MAC} detailing the histogram comparison for full data obtained through the simulation process but here accuracy of the histogram prediction achieved through a comparatively higher number of epochs.  
  
\section{Model Drawback}

Even though GCN models are achieving their accuracy within a reasonable period of time, the stability for providing good results can be degraded if we increase the number of hidden layers of graph convolutional neural networks beyond a certain number. This fact is very important if we consider very large circuits. But some researchers coming with new optimization methods to overcome the challenges faced by GCN.


\section{Future work}

\subsection{Individual flip-flop's FDR prediction }

\begin{figure}
    \centering
    \includegraphics[width=\linewidth]{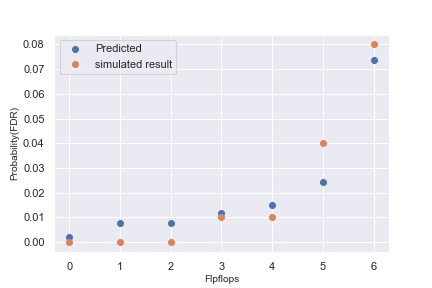}
    \caption{The trained flip-flops and it's predicted values without sorting}
    \label{figure_3_1}
\end{figure}

All the above analysis explaining an overall distribution and overall data envelope comparison (like histogram comparison), but the algorithm is not producing a comparison of individual FDR data prediction. This aim could be achievable. This clearly concluded from figure \ref{figure_3_1} because the individual predicted FDR of trained flip-flop sequence pretty well approximating to it's the original FDR. This example taken from the case of floating point adder circuit.  

Now after examining the trained sequence and it's predicted values from figure \ref{figure_3_1}, we can hope to extend this work with the training phase composed of a higher number of flip-flops for achieving more accurate values.  

\subsection{Classification or clustering of registers based on FDR}

It is already beginning to contemplate a future important application based on this model. It is the ability to do clustering registers based on the trained and predicted FDR. Once the model started to succeed in the prediction of individual FDR, then the classification aim will eventuate in reality.    

\section{Conclusion}

The works implemented in this paper depict the importance of modeling of FDR due to the soft error called SEU in microelectronic systems using a GCN network. An achieved goal by this model is the modeling of an arbitrary circuit with good accuracy and can predict the distribution of FDR derating factors. The detailed graphical comparison of predicted and stimulated FDR data for two completely different circuits, explicitly shows the prediction capability of the model. Future work for predicting more accurate individual flip-flop's FDR data going on, which may result in another dimension of applications including clustering of registers.   

\bibliographystyle{IEEEtran}
\bibliography{BIBLIOGRAPHY/IEEE_NASA_ESA_AHS_2019}

\end{document}